\documentclass[preprint,aps,superscriptaddress,nofootinbib,tightenlines]{revtex4}
\usepackage{epsfig}
\usepackage{bm}

\def\OMIT#1{{}}

\newcommand{\beq}{\begin{equation}}
\newcommand{\eeq}{\end{equation}}
\newcommand{\bea}{\begin{eqnarray}}
\newcommand{\eea}{\end{eqnarray}}

\newcommand{\benn}{\begin{displaymath}}
\newcommand{\eenn}{\end{displaymath}}

\begin{document}

\begin{figure}[!t]
\vskip -1.5cm
\leftline{
{\epsfxsize=1.8in \epsfbox{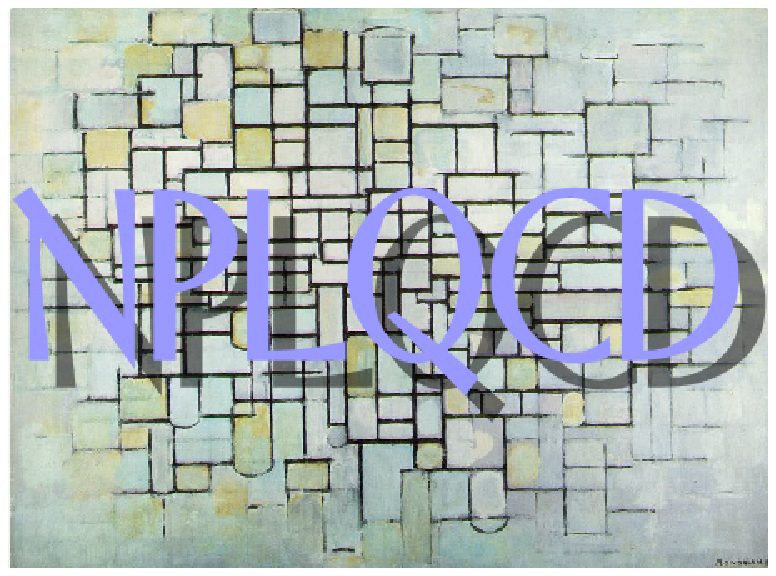}}}
\end{figure}

\preprint{\vbox{
\hbox{UNH-06-06}
\hbox{UMD-40762-369}
\hbox{JLAB-THY-06-514}
\hbox{UG-06-05}
\hbox{NT@UW-06-17}}}

\vphantom{}
\title{\bf \LARGE $\pi K$ Scattering in Full QCD with\\ Domain-Wall Valence Quarks}
\author{Silas R.~Beane}
\affiliation{Department of Physics, University of New Hampshire,
Durham, NH 03824-3568.}
\author{Paulo F.~Bedaque}
\affiliation{Department of Physics, University of Maryland, College Park, MD 20742-4111.}
\author{Thomas C.~Luu}
\affiliation{N Division, Lawrence Livermore National Laboratory, Livermore, CA 94551.}
\author{Kostas Orginos}
\affiliation{Department of Physics, College of William and Mary, Williamsburg,
  VA 23187-8795.}
\affiliation{Jefferson Laboratory, 12000 Jefferson Avenue, 
Newport News, VA 23606.}
\author{Elisabetta Pallante}
\affiliation{Institute for Theoretical Physics, University of Groningen,
Nijenborgh 4,\\ 9747 AG  Groningen, The Netherlands.}
\author{Assumpta Parre\~no}
\affiliation{Departament d'Estructura i Constituents de la Mat\`{e}ria,
Universitat de Barcelona, \\ E--08028 Barcelona, Spain.}
\author{Martin J.~Savage}
\affiliation{Department of Physics, University of Washington, 
Seattle, WA 98195-1560.\\
\qquad}
\collaboration{ NPLQCD Collaboration }
\noaffiliation
\vphantom{}
\vskip 0.8cm
\begin{abstract} 
\vskip 0.5cm
\noindent We calculate the $\pi^+K^+$ scattering length in
fully-dynamical lattice QCD with domain-wall valence quarks on MILC
lattices with rooted staggered sea-quarks at a lattice spacing of
$b=0.125~{\rm fm}$, lattice spatial size of $L=2.5~{\rm fm}$ and at pion
masses of $m_\pi\sim 290,350,490$ and $600~{\rm MeV}$.  The lattice
data, analyzed at next-to-leading order in chiral perturbation theory,
allows an extraction of the full $\pi K$ scattering amplitude at
threshold. Extrapolating to the physical point gives $m_\pi \ a_{3/2}
= -0.0574\pm 0.0016^{+0.0024}_{-0.0058}$ and $m_\pi \ a_{1/2} =
0.1725\pm 0.0017^{+0.0023}_{-0.0156}$ for the $I=3/2$ and $I=1/2$
scattering lengths, respectively, where the first error is statistical
and the second error is an estimate of the systematic due to
truncation of the chiral expansion.

\end{abstract}
\maketitle

\vfill\eject

\section{Introduction}
\noindent 

\noindent In hadronic atoms, nature has provided a relatively clean
environment in which to explore the low-energy interactions of charged
hadrons.  The electromagnetic interaction allows for
oppositely-charged, long-lived hadrons to form Coulomb bound states.
The locations of the energy-levels of these systems are perturbed by
the strong interactions, while the lifetimes of the ground states are
dictated by the strong interactions that couple the charged hadrons to
lighter neutral ones.

Theoretically, the simplest hadronic atom to understand is composed of
two pions: $\pi^+\pi^-$.  Precision experiments have been performed
---and are ongoing--- to measure the lifetimes and energy-levels of
such atoms~\cite{Adeva:2005pg}.  In the isospin limit, Bose statistics
dictates that two pions interacting in an s-wave can be in either an
isospin-0 or isospin-2 state.  By measuring the decay width and energy
levels of pionium, the $I=0$ and $I=2$ strong-interaction scattering
lengths can be isolated. While the difference between energy levels,
and hence deviations from the Coulomb spectrum, are relatively
straightforward to measure, it is somewhat more challenging to
determine the lifetime of these atoms.  Recently, the DIRAC
collaboration~\cite{Adeva:2005pg} at CERN has measured the lifetime to
be $\tau_{\pi^+\pi^-} = 2.91^{+0.49}_{-0.62}\times 10^{-15}~s$, with
the dominant decay mode being $\pi^+\pi^-\rightarrow 2\pi^0$.  On the
theoretical side, progress in lattice QCD has been quite rapid, with a
recent fully-dynamical calculation of the $I=2$ $\pi\pi$ scattering
length at pion masses between $m_\pi\sim 290~{\rm MeV}$ and $500~{\rm
MeV}$~\cite{Beane:2005rj}.  When combined with two-flavor chiral
perturbation theory, a prediction of the scattering length at the
physical point is found to have an uncertainty that is somewhat
smaller than that from experiment.  An up-to-date discussion of the
status of $\pi\pi$-interactions can be found in
Ref.~\cite{Caprini:2005an}.

Studying the low-energy interactions between kaons and pions with
$\pi^-K^+$ bound-states allows for an explicit exploration of the
three-flavor structure of low-energy hadronic interactions, an aspect
that is not directly probed in $\pi\pi$ scattering.  Experiments have
been proposed by the DIRAC collaboration~\cite{DIRACprops} to study
$\pi K$ atoms at CERN, J-PARC and GSI, the results of which would
provide direct measurements or constraints on combinations of the
scattering lengths.  In the isospin limit, there are two isospin
channels available to the $\pi K$ system, $I={1\over 2}$ and
$I={3\over 2}$.  The width of a $\pi^-K^+$ atom depends upon the
difference between scattering lengths in the two channels, $\Gamma\sim
(a_{1/2}-a_{3/2})^2$, (where $ a_{1/2}$ and $a_{3/2}$ are the
$I={1\over 2}$ and $I={3\over 2}$ scattering lengths, respectively)
while the shift of the ground-state depends upon a different
combination, $\Delta E_0\sim 2 a_{1/2}+a_{3/2}$.  Recently, the
Roy-Steiner equations (analyticity, unitarity and crossing-symmetry)
have been used to extrapolate high-energy $\pi K$ data down to
threshold~\cite{Buettiker:2003pp}, where it is found that
\begin{eqnarray}
m_\pi \left(a_{1/2}-a_{3/2}\right) & = & 0.269\pm 0.015
\ \ ,\ \ 
m_\pi \left(a_{1/2}+ 2 a_{3/2}\right) \ = \ 0.134\pm 0.037
\ \ \ ,
\label{eq:roy}
\end{eqnarray}
which can be decomposed to $m_\pi a_{1/2}=0.224\pm 0.022$ and $m_\pi
a_{3/2}=-0.0448\pm 0.0077$.  In addition, three-flavor chiral
perturbation theory ($\chi$PT) has been used to predict these
scattering lengths out to next-to-next-to-leading order (NNLO) in the
chiral expansion.  At NLO~\cite{Bernard:1990kw,Bernard:1990kx,Kubis:2001bx},
\begin{eqnarray}
m_\pi \left(a_{1/2}-a_{3/2}\right) & = & 0.238\pm 0.002
\ \ ,\ \ 
m_\pi \left(a_{1/2}+ 2 a_{3/2}\right) \ = \ 0.097\pm 0.047
\ \ \ ,
\label{eq:nlo}
\end{eqnarray}
while at NNLO~\cite{Bijnens:2004bu} $m_\pi a_{1/2}=0.220$ and $m_\pi
a_{3/2}= -0.047$~\footnote{ At tree-level,
Weinberg~\cite{Weinberg:1966kf} determined that $m_\pi a_{1/2}=0.137$
and $m_\pi a_{3/2}= -0.0687$.}.  One must be cautious in assessing the
uncertainties in these theoretical calculations, as one can only make
estimates based on power-counting for the contribution of higher-order
terms in the chiral expansion. There has been one determination of the
$\pi^+ K^+$ scattering length in quenched QCD~\cite{Miao:2004gy},
however the chiral extrapolation was restricted to tree level.

It is worth mentioning a novel motivation for accurate determinations
of meson-meson scattering from lattice QCD calculations. Recent work
has identified in a model-independent way the lowest-lying resonance
in QCD which appears in $\pi\pi$
scattering~\cite{Caprini:2005zr}. Crucial to this development has been
the accurate determination of the low-energy $\pi\pi$ scattering
amplitude, including the recent lattice QCD determination of the $I=2$
scattering length~\cite{Beane:2005rj}. A similar analysis has very
recently been carried out for $\pi K$ scattering in the
$I=\frac{1}{2}$ s-wave in order to determine the lowest-lying strange
resonance~\cite{Descotes-Genon:2006uk}. Improved accuracy in the
low-energy $\pi K$ scattering amplitude should be welcome to this
endeavor.

In this work we present the results of a fully-dynamical lattice QCD
calculation of $\pi^+K^+$ scattering. By calculating the $m_\pi$ and
$m_K$ dependence of the $\pi^+K^+$ ($I={3\over 2}$) scattering length,
we are able to provide a determination of both the $I={3\over 2}$ and
$I={1\over 2}$ scattering lengths at the physical point. We have
performed a hybrid mixed-action calculation with domain-wall valence
quarks tuned to the staggered sea-quark masses of the MILC
configurations.  As the computer resources do not presently exist to
perform such calculations at or very near the physical value of the
light-quark masses, these are performed at pion masses between $m_\pi
\sim 290~{\rm MeV}$ and $\sim 600~{\rm MeV}$.  These results are
combined with calculations in continuum three-flavor $\chi$PT to
extrapolate to the physical point.

\section{Finite-Volume Calculation of Scattering Amplitudes}
\label{sec:finvol}

\noindent The s-wave scattering amplitude for two particles below
inelastic thresholds can be determined using L\"uscher's
method~\cite{luscher_formula}, which entails a measurement of one or
more energy levels of the two-particle system in a finite volume.  For
two particles with masses $m_1$ and $m_2$ in an s-wave, with zero
total three momentum, and in a finite volume, the difference between
the energy levels and those of two non-interacting particles can be
related to the inverse scattering amplitude via the eigenvalue
equation~\cite{luscher_formula}
\begin{eqnarray}
p\cot\delta(p) \ =\ {1\over \pi L}\ {\bf S}\left(\,\frac{p L}{2\pi}\,\right)\ \ ,
\label{eq:energies}
\end{eqnarray}
where $\delta(p)$ is the elastic-scattering phase shift, and
the regulated three-dimensional sum is
\begin{eqnarray}
{\bf S}\left(\,{\eta}\, \right)\ \equiv \ \sum_{\bf j}^{ |{\bf j}|<\Lambda}
{1\over |{\bf j}|^2-\eta^2}\ -\  {4 \pi \Lambda}
\ \ \  .
\label{eq:Sdefined}
\end{eqnarray}
The sum in eq.~(\ref{eq:Sdefined})
is over all triplets of integers ${\bf j}$ such that $|{\bf j}| < \Lambda$ and the
limit $\Lambda\rightarrow\infty$ is implicit~\cite{Beane:2003da}. 
This definition is equivalent to the analytic continuation of zeta-functions presented 
by L\"uscher~\cite{luscher_formula}.
In eq.~(\ref{eq:energies}), $L$ is the length of the spatial dimension in a 
cubically-symmetric lattice.
The energy eigenvalue $E_n$ and its deviation from the sum of the  rest masses 
of the particle, $\Delta E_n$,
are related to the center-of-mass momentum $p_n$, a solution of
eq.~(\ref{eq:energies}), by
\begin{eqnarray}
\Delta E_n \ & \equiv & E_n\ -\  m_1 \ - \ m_2 \ =\ \sqrt{\ p_n^2\ +\ m_1^2\ } \ +\
\sqrt{\ p_n^2\ +\ m_2^2\ }
\ -\ m_1\  - \ m_2
\nonumber\\
& = & { p_n^2\over  2 \mu_{12}}\ +\ ...
\ \ \ ,
\label{eq:energieshift}
\end{eqnarray}
where $\mu_{12}$ is the reduced mass of the system.
In the absence of interactions between the particles,
$|p\cot\delta|=\infty$, and the energy levels occur at momenta ${\bf
p} =2\pi{\bf j}/L$, corresponding to single-particle modes in a cubic
cavity.  Expanding eq.~(\ref{eq:energies}) about zero momenta, $p\sim
0$, one obtains the familiar relation~\footnote{ We have chosen to use
the ``particle physics'' definition of the scattering length, as
opposed to the ``nuclear physics'' definition, which is opposite in
sign.}
\begin{eqnarray}
\Delta E_0 &  = &  -\frac{2\pi a}{\mu_{12}  L^3}
\left[\ 1\ +\  c_1 \frac{a}{L}\ +\  c_2 \left( \frac{a}{L} \right)^2 \ \right ]
\ +\ {\cal O}\left({1\over L^6}\right)
\ \ ,
\label{luscher_a}
\end{eqnarray}
with 
\begin{eqnarray}
c_1 & = & {1\over \pi}
\sum_{{\bf j}\ne {\bf 0}}^{ |{\bf j}|<\Lambda}
{1\over |{\bf j}|^2}\ -\   4 \Lambda \
\ =\ -2.837297
\ \ \ ,\ \ \
c_2\ =\ c_1^2 \ -\ {1\over \pi^2} \sum_{{\bf j}\ne {\bf 0}}
{1\over |{\bf j}|^4}
\ =\ 6.375183
\ \ ,
\end{eqnarray}
and $a$ is the scattering length, defined by
\begin{eqnarray}
a & = & \lim_{p\rightarrow 0}\frac{\tan\delta(p)}{p} 
\ \ \ .
\label{eq:scatt}
\end{eqnarray}
For the $I={3\over 2}$ $\pi K$ scattering length, $a_{3/2}$, that we compute in
this work, the difference between the exact solution to
eq.~(\ref{eq:energies}) and the approximate solution in
eq.~(\ref{luscher_a}) is much less than $1\%$.

\section{Details of the Lattice Calculation}
\label{sec:Latt}

\noindent 
Our computation uses the mixed-action lattice QCD scheme developed by
LHPC~\cite{Renner:2004ck,Edwards:2005kw} which places domain-wall
valence quarks from a smeared-source on $N_f=2+1$
asqtad-improved~\cite{Orginos:1999cr,Orginos:1998ue} MILC
configurations generated with rooted~\footnote{For recent discussions
of the ``legality'' of the mixed-action and rooting procedures, see
Ref.~\cite{Durr:2004ta,Creutz:2006ys,Bernard:2006vv,Durr:2006ze,Hasenfratz:2006nw,Bernard:2006ee,Shamir:2006nj}.}
staggered sea quarks~\cite{Bernard:2001av} that are hypercubic-smeared
(HYP-smeared)~\cite{Hasenfratz:2001hp,DeGrand:2002vu,DeGrand:2003in,Durr:2004as}.
In the generation of the MILC configurations, the strange-quark mass
was fixed near its physical value, $b m_s = 0.050$, (where
$b=0.125~{\rm fm}$ is the lattice spacing) determined by the mass of
hadrons containing strange quarks.  The two light quarks in the
configurations are degenerate (isospin-symmetric).  As was shown by
LHPC~\cite{Renner:2004ck,Edwards:2005kw}, HYP-smearing allows for a
significant reduction in the residual chiral symmetry breaking at a
moderate extent $L_s = 16$ of the extra dimension and domain-wall
height $M_5=1.7$.  Using Dirichlet boundary conditions we reduced the
original time extent of 64 down to 32. This allowed us to recycle
propagators computed for the nucleon structure function calculations
performed by LHPC. For bare domain-wall fermion masses we used the
tuned values that match the staggered Goldstone pion to few-percent
precision. For details of the matching see
Refs.~\cite{Renner:2004ck,Edwards:2005kw}.  The parameters used in the
propagator calculation are summarized in Table~\ref{tab:MILCcnfs}. All
propagator calculations were performed using the Chroma software
suite~\cite{Edwards:2004sx,McClendon:2001aa}.
\begin{table}[tbp]
 \caption{The parameters of the MILC gauge configurations and
   domain-wall propagators used in this work. The subscript $l$
   denotes light quark (up/down) where $s$ denotes the strange
   quark. The superscript $dwf$ denotes the bare quark mass for the
   domain wall fermion propagator calculation. The last column is
   number of configurations times number of sources per
   configuration.}
\label{tab:MILCcnfs}
\begin{ruledtabular}
\begin{tabular}{ccccccc}
 Ensemble        
&  $b m_l$ &  $b m_s$ & $b m^{dwf}_l$ & $ b m^{dwf}_s $ & $10^3 \times b
m_{res}$~\protect\footnote{Computed by the LHP collaboration.} & \# of propagators  \\
\hline 
2064f21b676m007m050 &  0.007 & 0.050 & 0.0081 & 0.081  & $1.604\pm 0.038$ & 468$\times$4 \\
2064f21b676m010m050 &  0.010 & 0.050 & 0.0138 & 0.081  & $1.552\pm 0.027$ & 658$\times$4 \\
2064f21b679m020m050 &  0.020 & 0.050 & 0.0313 & 0.081  & $1.239\pm 0.028$ & 486$\times$3 \\
2064f21b681m030m050 &  0.030 & 0.050 & 0.0478 & 0.081  & $0.982\pm 0.030$ & 564$\times$6 \\
\end{tabular}
\end{ruledtabular}
\end{table}

As it is the difference in the energy between interacting mesons and non-interacting mesons that
provides the scattering amplitude, we computed the one-pion correlation 
function $C_{\pi^+} (t)$, the one-kaon correlation function $C_{K^+} (t)$, and 
the kaon-pion correlation function $C_{\pi^+ K^+} (p,t)$, where $t$ denotes the 
number of time-slices between the hadronic-sink and the hadronic-source, and $p$ denotes the 
magnitude of the (equal and opposite) momentum of each meson.
The single-pion correlation function is 
\begin{eqnarray}
C_{\pi^+}(t) & = & \sum_{\bf x}
\langle \pi^-(t,{\bf x})\ \pi^+(0, {\bf 0})
\rangle
\ \ \ ,
\label{pi_correlator} 
\end{eqnarray}
where the summation over ${\bf x}$ corresponds to summing over all the spatial
lattice sites, thereby projecting onto 
the momentum ${\bf p}={\bf 0}$ state.
The single-kaon correlation function has  a similar form.
The $\pi^+K^+$ correlation function  that projects onto the s-wave state in the continuum limit is
\begin{eqnarray}
C_{\pi^+K^+}(p, t) & = & 
\sum_{|{\bf p}|=p}\ 
\sum_{\bf x , y}
e^{i{\bf p}\cdot({\bf x}-{\bf y})} 
\langle \pi^-(t,{\bf x})\ K^-(t, {\bf y})\ K^+(0, {\bf 0})\ \pi^+(0, {\bf 0})
\rangle
\ \ \ , 
\label{pipi_correlator} 
\end{eqnarray}
where, in eqs.~(\ref{pi_correlator}) and (\ref{pipi_correlator}),
$\pi^+(t,{\bf x}) = \bar u(t, {\bf x}) \gamma_5  d(t, {\bf x})$ is an interpolating field for the  
$\pi^+$, and 
$K^+(t,{\bf x}) = \bar u(t, {\bf x}) \gamma_5  s(t, {\bf x})$ is an interpolating field for the  
$K^+$.
In the relatively large lattice volumes that we are using, the energy
difference between the interacting and non-interacting two-meson states
is a small fraction of the total energy, which is dominated by the
masses of the mesons.  In order to extract this energy difference we
formed the ratio of correlation functions, $G_{\pi^+ K^+}(p, t)$, where
\begin{eqnarray}
G_{\pi^+ K^+}(p, t) & \equiv &
\frac{C_{\pi^+K^+}(p, t)}{C_{\pi^+}(t) C_{K^+}(t)} 
\ \rightarrow \ \sum_{n=0}^\infty\ {\cal A}_n\ e^{-\Delta E_n\ t} 
\  \ ,
\label{ratio_correlator} 
\end{eqnarray}
and the arrow becomes an equality in the limit of an infinite number of
gauge configurations.  In $G_{\pi^+ K^+}(p, t)$, some of the fluctuations
that contribute to both the one- and two-meson correlation functions
cancel, thereby improving the quality of the extraction of the energy
difference beyond what we are able to achieve from an analysis of the individual
correlation functions.

\section{Analysis and Chiral Extrapolation}
\noindent

\begin{figure}[!ht]
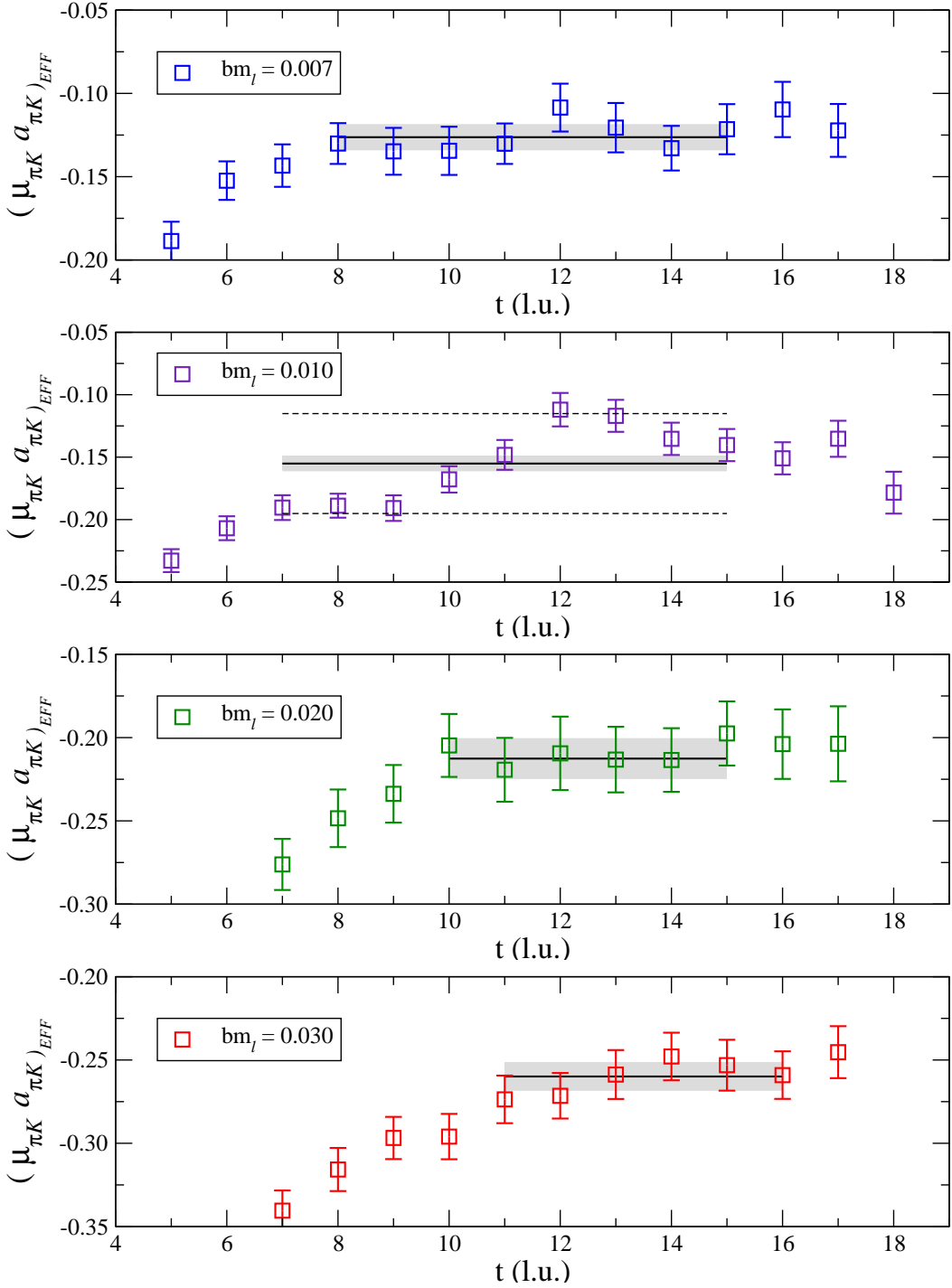

\centering                  
\includegraphics*[width=0.85\textwidth,viewport=2 5 700 240,clip]{EFFmua007.eps}
\hfill
\includegraphics*[width=0.85\textwidth,viewport=2 5 700 240,clip]{EFFmua010.eps}
\hfill
\includegraphics*[width=0.85\textwidth,viewport=2 5 700 240,clip]{EFFmua020.eps}
\hfill
\includegraphics*[width=0.85\textwidth,viewport=2 5 700 240,clip]{EFFmua030.eps}
\caption{\it 
The effective $\pi^+K^+$ scattering length times the reduced mass, $ \mu_{\pi K}\  a_{\pi^+K^+}(t)$
as a function of time slice arising from smeared sinks. The solid black lines and
shaded regions are fits with 1-$\sigma$ errors tabulated in Table~\ref{tab:LatResults}.
The dashed lines on the m010 ensemble plot are an estimate of a systematic error due
to fitting.}
\label{fig:SSKPplots}
\end{figure}

\noindent A convenient way to present the data is with ``effective
scattering length'' plots, simple variants of effective mass
plots.  The effective energy splitting is formed from the ratio of
correlation functions
\begin{eqnarray}
\Delta E_{\pi^+ K^+}(t) & = & \log\left({ G_{\pi^+ K^+}(0,t)\over  G_{\pi^+ K^+}(0,t+1)}\right)
\  \ ,
\label{eq:effene} 
\end{eqnarray}
which in the limit of an infinite number of gauge configurations would
become a constant at large times that is equal to the lowest energy of
the interacting kaon and pion in the volume.  At each time-slice,
$\Delta E_{\pi^+ K^+}(t)$ is inserted into eqs.(\ref{eq:energieshift}) and
(\ref{eq:energies}), or into eq.~(\ref{luscher_a}), to give a
scattering length at each time slice, $a_{\pi^+K^+}(t)$.  It turns out
to be more useful to consider the dimensionless quantity of the
reduced mass times the scattering length, $\mu_{\pi K}\ a_{\pi^+K^+}$,
in our analysis, where $\mu_{\pi K}(t)$, the ``effective reduced
mass'' is constructed from the effective mass of the single particle
correlators.  For each of the MILC ensembles that we analyze, the
effective scattering lengths are shown in fig.~\ref{fig:SSKPplots}.
We found  that the plateaus in $\mu_{\pi K}\  a_{\pi^+K^+}$ are better defined
for smeared sinks than those from point sinks. 

\begin{figure}[!ht]
\centering
\includegraphics*[width=0.85\textwidth]{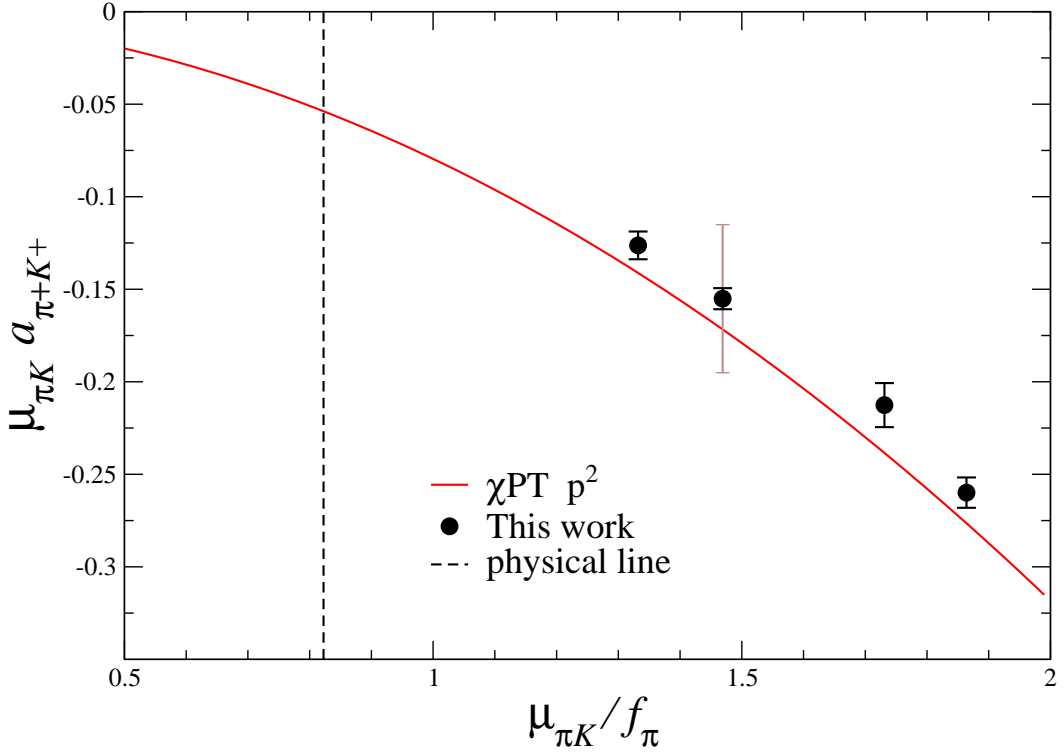}
\caption{$\mu_{\pi K} \ a_{\pi^+K^+}$ vs. $\mu_{\pi K}/f_\pi$.  The
data points are the results of this lattice calculation, while the
curve is the theoretical prediction at tree level in chiral
perturbation theory~\protect\cite{Weinberg:1966kf}.  The dark error
bar is statistical, while the lighter error bar corresponds to the
systematic error.  The vertical dashed line denotes the physical pion
and kaon masses.  }
\label{fig:muamu}
\end{figure}

The energy-shifts of the $\pi^+K^+$ groundstate can be extracted
directly from the effective scattering length function, or from
correlation functions, giving the same results.  A single exponential
function was fit by $\chi^2$-minimization to the correlation
functions, from which either the $\pi^+K^+$ energy or single particle
masses were determined.  The central value and uncertainty of each
parameter was determined by the jackknife procedure over the ensemble
of configurations.
\begin{table}[!h]
 \caption{Results from the lattice calculation. All errors are computed from
   jackknife.  The uncertainty associated with the m010 ensemble $\pi K$ energy
   shift and related quantities is dominated by the systematic error.
The fitting ranges are shown in the square brackets.
 }
\label{tab:LatResults}
\begin{ruledtabular}
\begin{tabular}{ccccccc}
 Ensemble        
&  $m_{\pi}/f_{\pi}$ & $m_K/f_\pi$ & $\mu_{\pi K}/f_\pi$ & $\delta E_{\pi K}$
(MeV)  &  
$\mu_{\pi K} \ a_{\pi^+K^+}$ & ${\Gamma}\times 10^3$\\
\hline 
m007 & 2.000(17)  & 3.980(25) & 1.332(10)& 11.89(81)\ [8-15]&  -0.1263(75)& -10.1(9)     \\
m010 & 2.337(11)  & 3.958(16) & 1.469(07)& 11.40(50)\ [7-15]&  -0.155(40)&  -8(3)\\
m020 & 3.059(12)  & 3.988(15) & 1.731(07)& 10.15(69)\ [10-15]&  -0.213(12)&  -5.59(34)\\
m030 & 3.484(10)  & 4.004(12) & 1.869(05)& 10.06(54)\ [11-16]&  -0.267(12)&  -4.29(15)
\end{tabular}
\end{ruledtabular}
\end{table}
The results of our lattice calculation of the 
decay constants, meson masses, $\pi^+K^+$ energy shifts and scattering lengths 
are tabulated in Table~\ref{tab:LatResults}.
The scattering lengths as a function of reduced mass are shown in fig.~\ref{fig:muamu}.

As can be seen from fig.~\ref{fig:SSKPplots}, there is a large
systematic error associated with the m010 ensemble. There would appear
to be two distinct plateaus. Rather than fitting to one of the
plateaus, we chose to fit over a large range (7-15) which includes
both plateaus and then assigned a systematic error which encompasses
minima and maxima over the fit range as indicated by the effective
scattering length plot. More statistics will have to be acquired on
this ensemble before any conclusions can be drawn about this correlator.

In SU(3) chiral perturbation
theory~\cite{Gasser:1984gg,Gasser:1983yg,Gasser:1983ky} at NLO, the
expansion of the crossing even ($a^+$) and crossing odd ($a^-$)
scattering length times the reduced mass is known to
be~\cite{Bernard:1990kw,Bernard:1990kx,Kubis:2001bx}
\begin{eqnarray}
\mu_{\pi K} a^- \!&=&\! \frac{\mu_{\pi K}^2}{4\pi f_\pi^2} \Biggl\{ 1 
\,+\,
2\frac{m_\pi^2}{f_\pi^2} \,\Biggl[ \, 8 L_5(\lambda)
\label{a0minus} \\
\!&-&\!  \frac{1}{16\pi^2} 
\Biggl\{
\frac{8m_K^2-5m_\pi^2}{2(m_K^2 - m_\pi^2)}  \log \frac{m_\pi}{\lambda}
-\frac{23m_K^2}{9(m_K^2 - m_\pi^2)} \log \frac{m_K}{\lambda} 
+ \frac{28m_K^2 - 9m_\pi^2}{18(m_K^2 - m_\pi^2)}  \log \frac{m_\eta}{\lambda}
\nonumber\\ && 
\quad +\frac{4m_K}{9m_\pi}
\frac{\sqrt{(m_K - m_\pi)(2m_K + m_\pi)}}{m_K+m_\pi}
  \arctan \Biggl( \frac{2(m_K + m_\pi)}{m_K - 2m_\pi} \sqrt{\frac{m_K - m_\pi}{2m_K + m_\pi}} \Biggr)
\nonumber\\ && \quad 
-\frac{4m_K}{9m_\pi}
\frac{\sqrt{(m_K + m_\pi)(2m_K - m_\pi)}}{m_K-m_\pi} 
  \arctan \Biggl( \frac{2(m_K - m_\pi)}{m_K + 2m_\pi} \sqrt{\frac{m_K + m_\pi}{2m_K - m_\pi}} \Biggr)
\Biggr\} \Biggr] \Biggr\}
~, \nonumber \\
& \equiv & 
\frac{\mu_{\pi K}^2}{4\pi f_\pi^2} 
\Biggl[ 1 \,+\, 16\frac{m_\pi^2}{f_\pi^2} L_5\ +\ 
\chi^{(NLO,-)}\Biggr]
~, \nonumber \\
\mu_{\pi K} a^+ \!&=&\! \frac{\mu_{\pi K}^2 m_K m_\pi}{2\pi f_\pi^4} \Biggl[
16\  L_{\pi K}(\lambda)
\label{a0plus} \\ 
\!&+&\! \frac{1}{16\pi^2} \Biggl\{
\frac{11m_\pi^2}{2(m_K^2-m_\pi^2)} \log\frac{m_\pi}{\lambda}
-\frac{67m_K^2-8m_\pi^2}{9(m_K^2-m_\pi^2)} \log\frac{m_K}{\lambda}
+\frac{24m_K^2-5m_\pi^2}{18(m_K^2-m_\pi^2)} \log\frac{m_\eta}{\lambda} 
\nonumber\\ && 
\quad -\frac{4}{9}
\frac{\sqrt{(m_K - m_\pi)(2m_K + m_\pi)}}{m_K+m_\pi}
  \arctan \Biggl( \frac{2(m_K + m_\pi)}{m_K - 2m_\pi} \sqrt{\frac{m_K - m_\pi}{2m_K + m_\pi}} \Biggr)
\nonumber\\ && \quad 
-\frac{4}{9}
\frac{\sqrt{(m_K + m_\pi)(2m_K - m_\pi)}}{m_K-m_\pi} 
  \arctan \Biggl( \frac{2(m_K - m_\pi)}{m_K + 2m_\pi} \sqrt{\frac{m_K + m_\pi}{2m_K - m_\pi}} \Biggr)
+\frac{43}{9}
\Biggr\} \Biggr]~, 
\nonumber\\
 & \equiv & \frac{\mu_{\pi K}^2 m_K m_\pi}{2\pi f_\pi^4} \Biggl[
16\  L_{\pi K}
\ +\  \chi^{(NLO,+)}\Biggr]
\nonumber
\end{eqnarray}
where the counterterm $L_{\pi K}(\lambda)$ is a renormalization scale,
$\lambda$, dependent linear combination of the Gasser-Leutwyler
counterterms
\begin{eqnarray}
L_{\pi K} &\equiv& 2L_1 + 2L_2 + L_3 - 2L_4 - \frac{L_5}{2} + 2L_6 + L_8 
\ .
\end{eqnarray}
It is important to note that the expressions in eqs.~(\ref{a0minus}) and
(\ref{a0plus}) are written in terms of the full $f_\pi$, and not the chiral
limit value.
The functions $\chi^{(NLO,+)}\left(m_\pi/\lambda,m_K/\lambda,m_\eta/\lambda\right)$ 
and $\chi^{(NLO,-)}\left(m_\pi/\lambda,m_K/\lambda,m_\eta/\lambda\right)$
clearly depend upon the renormalization scale $\lambda$.
In the analysis that follows, it was found to be convenient to normalize the
meson masses to $f_\pi$, and therefore we can choose the renormalization scale
to be $\lambda=f_\pi^{\rm phys}$, and use the values of $m_\pi/f_\pi$ and
$m_K/f_\pi$ in Table~\ref{tab:LatResults} directly.  Deviations between the
$\lambda=f_\pi$ calculated on each lattice and $\lambda=f_\pi^{\rm phys}$ are
higher order in the chiral expansion.

\begin{figure}[!ht]
\centering
\includegraphics*[width=0.85\textwidth]{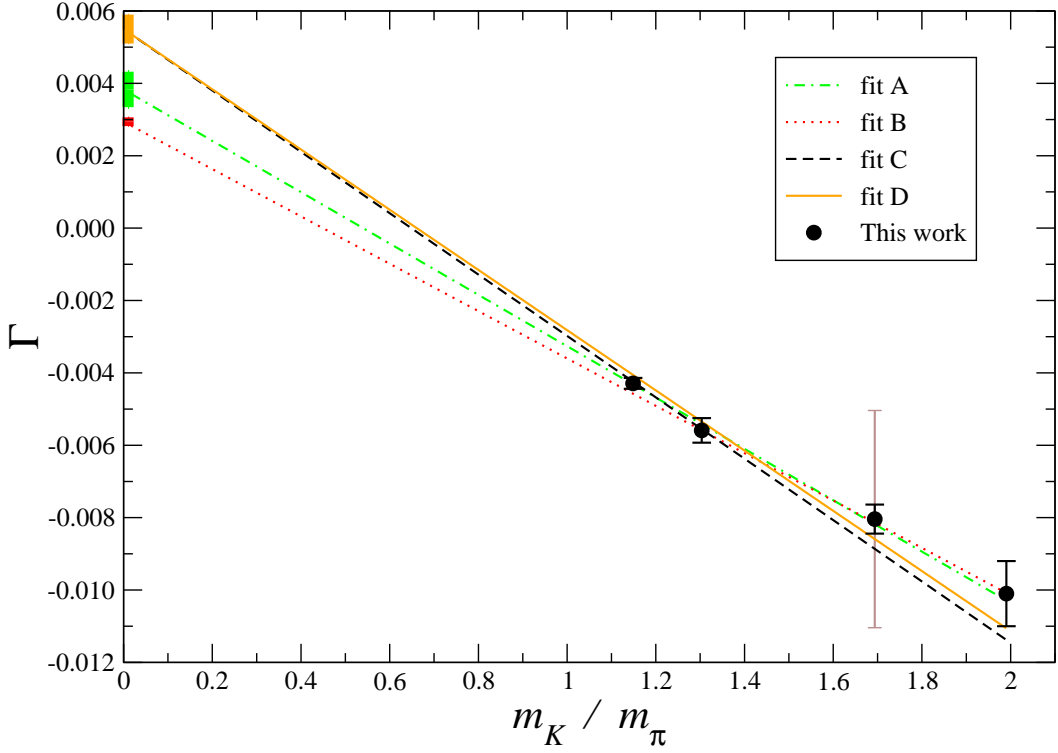}
\caption{${\Gamma}$ vs. $m_K/m_\pi$.
The dark error bar on the data points is statistical, while 
the lighter error bar corresponds to the systematic error.
The lines correspond to the four linear fits (A,B,C,D). The bars on
the y axis represent the 1-$\sigma$ errors in the determinations of $L_5=\Gamma(m_K/m_\pi =0)$
as given in Table~\ref{tab:FitResultsNNLO}. (At 95\% confidence level,
these determinations are in agreement.)}
\label{fig:G}
\end{figure}

The $I={1\over 2}$ and $I={3\over 2}$ scattering lengths are related to those in
eqs.~(\ref{a0minus}) and (\ref{a0plus}) by 
\begin{eqnarray}
a_{1/2} & = & a^+ \ +\ 2 a^- 
\nonumber\\
a_{3/2} & = & a^+ \ -\ a^-\ =\ a_{\pi^+K^+} 
\ \  .
\end{eqnarray}
It is convenient to define the function ${\Gamma}$ via a subtraction of the
tree-level and one-loop contributions in order to 
isolate the counterterms, 
\begin{eqnarray}
{\Gamma}\left({m_\pi\over f_\pi},{m_K\over f_\pi}\right)  & \equiv & 
-
{f_\pi^2\over 16 m_\pi^2}
\left( {4\pi f_\pi^2\over\mu_{\pi K}^2} \left[ \mu_{\pi K}\ a_{\pi^+K^+} \right]
 + 1 + \chi^{(NLO,-)} - 2 {m_K m_\pi\over f_\pi^2} \chi^{(NLO,+)}
\right) ,
\label{eq:Gdef}
\end{eqnarray}
where we use the Gell-Mann--Okubo mass-relation among the mesons to
determine the $\eta$-mass, which we do not measure in this
lattice calculation.  At NLO this becomes
\begin{eqnarray}
{\Gamma} & = & 
L_5(f_\pi^{\rm phys})\ -\ 2\ {m_K\over m_\pi}\  L_{\pi K}(f_\pi^{\rm phys})
\ .
\label{eq:Gdefrhs}
\end{eqnarray}
It is clear that the dependence of ${\Gamma}$ on $m_\pi$ and $m_K$ 
determines $L_5$ and $L_{\pi K}$ and, in turn, allows an
extraction of $a_{3/2}$ and $a_{1/2}$. The numerical values of
${\Gamma}$ and their jackknife errors calculated on each ensemble of
lattices are given in Table~\ref{tab:LatResults}, and are plotted in
fig.~\ref{fig:G}.
By fitting a straight line to the values of ${\Gamma}$ as a function of
$m_k/m_\pi$ the counterterms $L_5$ and $L_{\pi K}$ (renormalized at 
$f_\pi^{\rm phys}$) can be determined.  

Ideally, one would fit to lattice data at the lightest accessible
values of the quark masses in order to ensure convergence of the
chiral expansion.  While we only have four different quark masses in
our data set, with pion masses ranging from $m_\pi\sim 290~{\rm MeV}$
to $600~{\rm MeV}$, fitting all four data sets and then ``pruning''
the heaviest data set and refitting provides a useful measure of the
convergence of the chiral expansion. Hence, in
``fit A'', we fit the data from all four lattice ensembles (m007,
m010, m020 and m030), while in ``fit B'', we fit the data from the
lightest three lattice ensembles (m007, m010 and m020). 
\begin{table}[!ht]
 \caption{Results of the NLO fits.  The values of $m_\pi a_{3/2}$ and $m_\pi a_{1/2}$ 
correspond to their extrapolated values at the physical point, where the error ellipses in the $L_5$-$L_{\pi K}$ plane
have been explored at 68\% confidence level (see fig.~\protect\ref{fig:ellipsesL5LpiK}).
}
\label{tab:FitResultsNNLO}
\begin{ruledtabular}
\begin{tabular}{cccccc}
FIT &  ${L}_5\times 10^3$ & $L_{\pi K}\times 10^3$ &  $m_\pi  a_{3/2}$ &  $m_\pi  a_{1/2}$  &$\chi^2$/dof \\
\hline
A & $3.83\pm0.49$  & $3.55\pm0.20$  & $-0.0607\pm0.0025$ & $0.1631\pm0.0062$  & $0.17$  \\
B & $2.94\pm0.07$  & $3.27\pm0.02$  & $-0.0620\pm0.0004$ & $0.1585\pm0.0011$  & $0.001$  \\ 
C & $\ \ 5.65\pm0.02^{+0.18}_{-0.54}$~\protect\footnote{Input from $f_K/ f_\pi$~\protect\cite{Beane:2006kx}.}  & $4.24 \pm 0.17$  & $-0.0567\pm0.0017$ & $0.1731\pm0.0017$ & $0.84$  \\ 
D & $\ \ {5.65\pm0.02^{+0.18}_{-0.54}}^{\ \it a}$ & $4.16 \pm0.18$  & $-0.0574\pm0.0016$ & $0.1725\pm0.0017$ & $0.90$  \\ 
\end{tabular}
\end{ruledtabular}
\end{table}

\begin{figure}[!ht]
\centering
\includegraphics*[width=0.85\textwidth]{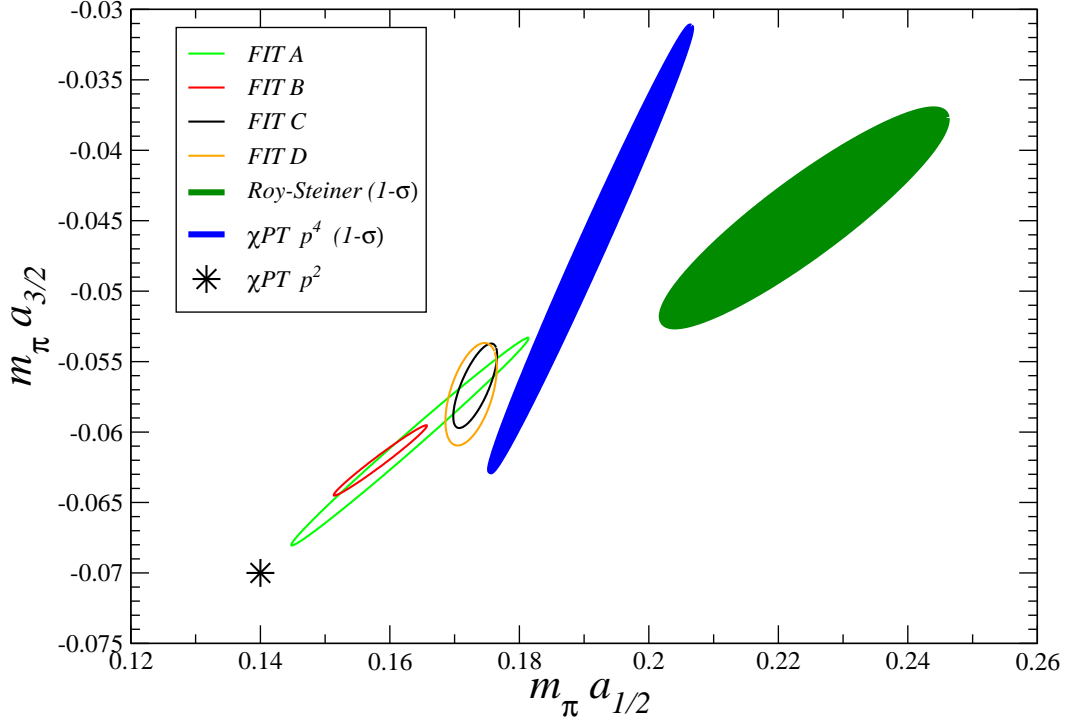}
\caption{Error ellipses for the four fits (A,B,C,D) at 95\% confidence
level. (Note that these results are derived from lattice data on a
single lattice spacing of $b=0.125~{\rm fm}$.).  The star corresponds
to the current-algebra predictions ($\chi$PT $p^2$) from
Ref.~\protect\cite{Weinberg:1966kf}. We also display 1-$\sigma$ error
ellipses from a $\chi$PT analysis at NLO~\protect\cite{Bernard:1990kw}
(denoted $\chi$PT $p^4$) and from a fit using the Roy-Steiner
equations~\protect\cite{Buettiker:2003pp}.}
\label{fig:ellipses}
\end{figure}

With the limited data set presently at our disposal, it is not
practical to fit to the NNLO expression~\cite{Bijnens:2004bu} for the
scattering length.  However, it is important to estimate the
uncertainty in the values of the scattering lengths extrapolated to
the physical point that is introduced by the truncation of the chiral
expansion at NLO.  In our work on $f_K/f_\pi$~\cite{Beane:2006kx} we
extracted a value of $L_5$ as it is the only NLO counterterm that
contributes.  The numerical value obtained is only perturbatively
close to its true value, as it is contaminated by higher-order
contributions.  Therefore, by fixing the $L_5$ that appears in
eq.~(\ref{eq:Gdefrhs}) to the value of $L_5$ extracted from
$f_K/f_\pi$, an estimate of the uncertainty in both $L_{\pi K}$ and in
the extrapolated values of the scattering lengths due to the
truncation of the chiral expansion can be estimated.  Specifically, we
sampled $L_5$ from a Gaussian distribution for a range of $f_K/f_\pi$
values~\cite{Beane:2006kx} and then fit $L_{\pi K}$ using
$\chi^2$-minimization. We then generated a value of $L_{\pi K}$ from a
normal distribution formed from its mean and standard error.
\begin{figure}[!ht]
\centering
\includegraphics*[width=0.85\textwidth]{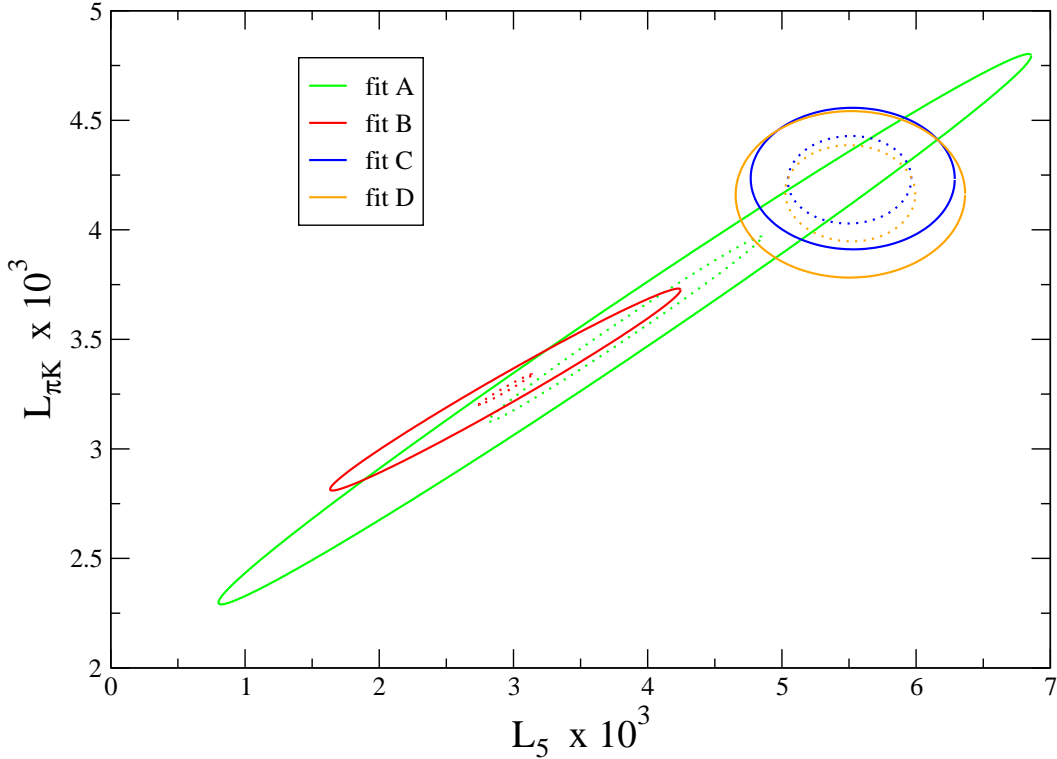}
\caption{Error ellipses in the $L_5$-$L_{\pi K}$ plane for the four fits (A,B,C,D) at 68\% (dotted lines) and 95\% (solid lines)
confidence level.}
\label{fig:ellipsesL5LpiK}
\end{figure}
This fit is denoted ``fit C'', and the same fit but with the m030 data
pruned is denoted ``fit D''. The results of the four fits are given in
Table~\ref{tab:FitResultsNNLO} and plotted in fig.~\ref{fig:G}. These
fits lead to an extraction of
\begin{eqnarray}
L_{\pi K} & = & 4.16 \pm 0.18 ^{+0.26}_{-0.91}
 \ \ \ ,
\end{eqnarray}
and a prediction of the scattering lengths extrapolated to the physical point of
\begin{eqnarray}
m_\pi \  a_{3/2} & = & -0.0574\pm 0.0016^{+0.0024}_{-0.0058} \nonumber \\ 
m_\pi \  a_{1/2} & = &\ \ 0.1725\pm 0.0017^{+0.0023}_{-0.0156} \ .
\end{eqnarray}
We have chosen to take the central values and statistical errors from
fit D and have set the systematic error due to truncation of the
chiral expansion by taking the range of the various quantities allowed
by the four fits, including statistical and systematic errors. 
In fig.~\ref{fig:ellipsesL5LpiK} we plot the 68\%
and 95\% confidence-level error ellipses for the four fits given in
Table~\ref{tab:FitResultsNNLO} in the $L_5$-$L_{\pi K}$ plane.  In
fig.~\ref{fig:ellipses} we plot the 95\% confidence-level error
ellipses associated with the four fits in the $m_\pi \ a_{1/2}$-$m_\pi
\ a_{3/2}$ plane~\footnote{In Mathematica format, the 95\% confidence-level error ellipses are:\\
fit A: Ellipsoid[\{0.1631,-0.0607\},\{0.0197,0.0007\},\{\{0.9283,0.3719\},\{-0.3719,0.9283\}\}]\\
fit B: Ellipsoid[\{0.1585,-0.0620\},\{0.0076,0.0004\},\{\{0.9461,0.3239\},\{-0.3239,0.9461\}\}]\\
fit C: Ellipsoid[\{0.1731,-0.0567\},\{0.0042,0.0016\},\{\{0.7534,0.6576\},\{-0.6576,0.7534\}\}]\\
fit D: Ellipsoid[\{0.1725,-0.0574\},\{0.0046,0.0027\},\{\{0.7881, 0.6156\},\{-0.6156,0.7881\}\}].
}.  For purposes of comparison we have included the
current-algebra point~\cite{Weinberg:1966kf} on the plot as well as
1-$\sigma$ error ellipses from analyses based on fitting
experimental data using $\chi$PT at NLO~\cite{Bernard:1990kw} and
using Roy-Steiner equations~\cite{Buettiker:2003pp}. As 1-$\sigma$
error ellipses correspond to 39\% confidence level, one should be
careful in finding discrepancy between the various determinations of
the scattering lengths. It would be interesting to see the NLO
$\chi$PT and Roy-Steiner error ellipses at higher confidence levels.

Given how well our lattice data fit the NLO continuum $\chi$PT
formulas, it would seem that the $O(b^2)$ discretization errors are
comparable or smaller than the systematic error due to omitted
$O(m_q^3)$ effects in the chiral expansion. However, one should keep
in mind that our determinations of, for instance, the low-energy
constants $L_5$ and $L_{\pi K}$ are subject to $O(b^2)$ shifts.  In
contrast with the $\pi^+\pi^+$ and $K^+K^+$ scattering lengths, the
mixed-action quantity $\Delta_{\rm Mix}$ makes an explicit
contribution to the $K^+\pi^+$ scattering length~\cite{Chen:2005ab,donal}.
While this adds an additional unknown contribution to this process, a
mixed-action $\chi$PT analysis of $\pi K$ scattering, including
lattice data from the fine MILC lattices ($b\sim0.09~{\rm fm}$), will
be able to address this source of systematic error quantitatively. We
continue to search for the computational resources to accomplish this
task.

\section{Conclusions}
\label{sec:resdisc}

\noindent In this paper we have computed the $\pi^+K^+$ scattering
length in fully-dynamical lattice QCD at pion masses ranging between
$m_\pi\sim 290~{\rm MeV}$ and $600~{\rm MeV}$.  We have used the
continuum expressions for the scattering lengths in SU(3) chiral
perturbation theory, together with lattice data for $f_K/f_\pi$, to
predict the physical $I=3/2$ and $I=1/2$ $\pi K$ scattering lengths
with unprecedented accuracy. Naively one would expect that $\pi^+K^+$
scattering would give information about $I=3/2$ scattering only.
However, the lattice data, when combined with 
chiral perturbation theory, implies a constraint on $I=1/2$
scattering as well. We anticipate that with improved statistics,
together with calculations on lattices with smaller lattice spacings,
the theoretically-predicted regions for $m_\pi a_{3/2}$ and $m_\pi
a_{1/2}$ can be further reduced beyond those shown in
fig.~\ref{fig:ellipses}. These regions can then be compared with the
expected measurements from $K^+\pi^-$ atoms, to provide an exciting
test of hadronic theory.

\acknowledgments

\noindent We thank Andre Walker-Loud and Donal O'Connell for useful
conversations and R.~Edwards for help with the QDP++/Chroma
programming environment~\cite{Edwards:2004sx} with which the
calculations discussed here were performed. The computations for this
work were performed at Jefferson Lab (JLab), Fermilab and Lawrence
Livermore National Laboratory. We are indebted to the MILC and the LHP
collaborations for use of their configurations and propagators,
respectively.  The work of MJS was supported in part by the
U.S.~Dept.~of Energy under Grant No.~DE-FG03-97ER4014. The work of KO
was supported in part by the U.S.~Dept.~of Energy contract
No.~DE-AC05-06OR23177 (JSA) and contract No.~DE-AC05-84150 (SURA).
The work of PFB was supported in part by the U.S.~Dept.~of Energy grant
No.~ER-40762-365. The work of SRB was supported in part by the
National Science Foundation under grant No.~PHY-0400231.  Part of this
work was performed under the auspices of the US DOE by the University
of California, Lawrence Livermore National Laboratory under Contract
No. W-7405-Eng-48. AP is supported by the Ministerio de Educaci\'on y
Ciencia (Spain) under contract No.~FIS2005-03142 and by the
Generalitat de Catalunya under contract No.~2005SGR-00343.

\end{document}